\documentstyle[12pt]{article}
\title{{\Large Electroweak Corrections to the Deacy $H^+\rightarrow
W^+h$\\ in the Minimal Supersymmetric Model }}
\author{  Ya Sheng Yang\thanks{E-mail: ylsdd@ibm320h.phy.pku.edu.cn} and
  Chong Sheng Li\thanks{E-mail: csli@ibm320h.phy.pku.edu.cn}\\
  \small{ Department of Physics, Peking University, Beijing 100871, P.R. China.}}

\begin{document}
\pagestyle{plain} \setcounter{page}{1} \baselineskip=0.3in

\maketitle

\vspace{.2in}
\begin{footnotesize}
\begin{center}\begin{minipage}{5in}
\baselineskip=0.25in
\begin{center} ABSTRACT \end{center}

We calculate the $O(\alpha_{ew}m_{t(b)}^{2}/m_{W}^{2})$ and
$O(\alpha_{ew} m_{t(b)}^4/m_W^4)$ supersymmetric electroweak
corrections to the process $H^+\rightarrow W^+h$ in the Minimal
Supersymmetric Model. These corrections arise from the virtual
effects of the third family (top and bottom) quaurks and squarks
(top-squark and bottom-squark). We find that for $m_{H^+}>200$GeV
at low $\tan\beta(\leq 3)$, the corrections can increase the
tree-level decay widths and the branching radios more than $20\%$
and $40\%$, respectively.

\end{minipage}\end{center}
\vspace{3.5cm}

\end{footnotesize}



 \eject \baselineskip=0.3in
\begin{center} {\Large 1. Introduction}\end{center}

The minimal supersymmeytic standard model(MSSM) takes the minimal
Higgs structure of two doubles[1], which predicts the existence of
three neutral and two charged Higgs bosons $h,H,A,$ and $H^{\pm}$.
When the Higgs boson of the Standard Model(SM) has a mass below
130-140 Gev and the h boson of the MSSM is in the decoupling limit
(which means that $H^+$ is too heavy anyway to be possibly
produced), the lightest neutral Higgs boson may be difficult to
distinguish from the neutral Higgs boson of the standard
model(SM). But charged Higgs bosons carry a distinctive signature
of the Higgs sector in the MSSM. Therefore, the search for charged
Higgs bosons is very important for probing the Higgs sector of the
MSSM and, therefore, will be one of the prime objectives of the
CERN Large Hadron Collider(LHC). In fact, in much of the parameter
space preferred by MSSM, namely $M_{H^{\pm}}> M_{W^{\pm}}$ and $1
< \tan\beta < m_t/m_b$[2,3], the LHC will provide the greatest
opportunity for the dicovery of $H^{\pm}$ particles. Previous
studies [4,5] have shown that the dominante decay modes of the
charged Higgs boson are $H^+\rightarrow t + \bar{b}$ with
$m_{H^{\pm}}>m_t + m_b$ and $H^+\rightarrow \tau ^+\nu$. However,
recent analyses[6,7] indicate that the altermative decay channel
$H^+\rightarrow W ^+h^0$ could be very important. In fact, its
branching ratio can be rather large, competing with the top-bottom
decay mode and overwhelming  the tau-neutrino one for
$M_{H^{\pm}}\geq m_t$ at low $\tan\beta$[7].  Thus an more
accurate calculation of the decay mechanisms is also necessary to
provide a solid basis for experimental analysis of observing
$H^+\rightarrow W ^+h^0$ at the LHC. In Ref.[6], R. Santos et al.
calculated the top quark loops corrections to the decay width of
the process $H^+\rightarrow W ^+h^0$ in the framework of the
two-Higgs doublet model for some reasonable choice of the free
parameters and found that such corrections can be as large as
$40\%$. In this letter, we present the calculation of the
$O(\alpha_{ew}m_{t(b)}^{2}/m_{W}^{2})$ and $O(\alpha_{ew}
m_{t(b)}^4/m_W^4)$ supersymmetric(SUSY) electroweak(EW)
corrections to this process in the MSSM. These corrections arise
from the virtual effects of the third family (top and bottom)
quaurks and squarks (top-squark and bottom-squark). We will give
attention manily to these corrections in parameter range
$\tan\beta \leq 3$ and $m_{H^+}> 200$ GeV, where the
$H^+\rightarrow W ^+h^0$ decay rate is significant.

\begin{center} {\Large 2. Calculations}\end{center}

Feynman diagrams contributing to supersymmetric electroweak
corrections to $H^+\rightarrow W^+h$ are shown in Fig.1 and 2. We
carried out the calculation in the t'Hooft-Feynman gauge and used
dimensional reduction, which preserves supersymmetry, for
regularization of the ultraviolet divergences in the virtual loop
corrections using the on-mass-shell renormalization scheme[8], in
which the fine-structure constant $\alpha_{ew}$ and physical
masses are chosen to be the renormalized parameters, and finite
parts of the counterterms are fixed by the renormalization
conditions. The coupling constant $g$ is related to the input
parameters $e$, $m_W,$ and $m_Z$ via $g^2= e^2/s_w^2$ and
$s_w^2=1-m_W^2/m_Z^2$. As far as the parameters $\beta$ and
$\alpha$, for the MSSM we are considering, they have to be
renormalized, too. In the MSSM they are not independent.
Nevertheless, we follow the approach of Mendez and Pomarol[9] in
which they consider them as independent renormalized parameters
and fixed the corresponding renormalization constants by a
renormalization condition that the on-mass-shell $H^+\bar l \nu_l$
and $h \bar l l$ couplings keep the forms of Eq.(3) of Ref.[9] to
all order of perturbation theory.

The relevant renormalization constants are defined as
\begin{eqnarray}
& & m_{W0}^2 =m_W^2 +\delta m_W^2,\ \ \ m_{Z0}^2 =m_Z^2 +\delta
m_Z^2, \nonumber
\\ & & \tan\beta_0 =(1+\delta Z_\beta)\tan\beta, \nonumber
\\ & & \sin\alpha_0 =(1+\delta Z_\alpha)\sin\alpha, \nonumber
\\ & & W_0^{\pm \mu} =(1+\delta Z_W)^{1/2}W^{\pm\mu}
+iZ_{H^{\pm}W^{\pm}}^{1/2}\partial^\mu H^\mp, \nonumber
\\ & & H_0^\pm =(1+\delta Z_{H^\pm})^{1/2}H^\pm, \nonumber
\\ & & H_0 =(1+\delta Z_H)^{1/2}H +Z_{Hh}^{1/2}h, \nonumber
\\ & & h_0 =(1+\delta Z_h)^{1/2}h +Z_{hH}^{1/2}H.
\end{eqnarray}

Taking into account the $O(\alpha_{ew}m_{t(b)}^{2}/m_{W}^{2})$ and
$O(\alpha_{ew}m_{t(b)}^{4}/m_{W}^{4})$ SUSY EW corrections, the
renormalized amplitude for $H^+\rightarrow W^+h$ can be written as
\begin{eqnarray}
M_{ren} = M_0+\delta M,
\end{eqnarray}
where $M_0$ is the tree-level amplitude arising from Fig.1$(a)$,
which is given by
\begin{eqnarray}
M_0=-\frac{ig}{2}\cos(\beta-\alpha) p^\mu \varepsilon_\mu(k).
\end{eqnarray}
Here $p$ is the momentum of the incoming charged higges Boson, and
$k$ is the momentum of the outgoing $W^+$ Boson. $\delta M$
represents the SUSY EW corrections, which is given by
\begin{eqnarray}
&&\delta M =-\frac{ig}{2}\cos(\beta-\alpha)p^\mu
\varepsilon_\mu(k)[\frac{\delta
g}{g}-\tan(\beta-\alpha)(\cos\beta\sin\beta\delta
Z_\beta-\tan\alpha\delta Z_\alpha)\nonumber\\
&&+\frac{1}{2}(\delta Z_{H^+}+\delta Z_h+\delta
Z_W)-\tan(\beta-\alpha)\delta Z_{Hh}^{1/2}]\nonumber\\ &&
+igm_W\sin(\beta-\alpha)Z_{HW}^{1/2}p^\mu+i\sum_{j=1}^4\Lambda_j
p^\mu\varepsilon_\mu(k),
\end{eqnarray}
with
\begin{eqnarray}
&& \frac{\delta g}{g}= \frac{\delta e}{e}+\frac{1}{2}\frac{\delta
m_Z^2}{m_Z^2}-\frac{1}{2}\frac{\delta m_Z^2-\delta
m_W^2}{m_Z^2-m_W^2}\nonumber,
\\&& \delta Z_\beta =-\frac{\delta g}{g}+ \frac{1}{2}\frac{\delta m_W^2}{m_W^2}
-\frac{1}{2}\delta Z_{H^+} - \frac{m_W}{\tan \beta}
Z_{HW}^{1/2}\nonumber,
\\&& \delta Z_\alpha = -\frac{\delta g}{g}+ \frac{1}{2}\frac{\delta
m_W^2}{m_W^2} -\frac{1}{2}\delta Z_h - \cot \alpha Z_{Hh}^{1/2} -
\sin^2 \beta \delta Z_\beta.
\end{eqnarray}
Here $\Lambda_j(j=1$-$4)$ are the vertex form factors coming from
Fig.1$(b)$-$(e)$. The $\delta e/e$ appearing in Eq.(5) does not
contain the $O(\alpha_{ew} m_{t(b)}^2/m_W^2)$ corrections and
needs not be considered in our calculations.

Calculating the self-energy diagrams in Fig.2, we can get the
explicit expressions of all the renormalization constants as
following:
\begin{eqnarray}
&& \delta m_W^2 =\frac{g^2}{16\pi^2}\{(m_b^2 -m_t^2)(1
  +\frac{m_b^2 -m_t^2 -2m_W^2}{2m_W^2}B_0^{0bt})-2m_t^2B_0^{0tt}
  \nonumber\\ && \hspace{1.0cm} -\frac{1}{2m_W^2}[(m_b^2 -m_t^2)^2 +(m_b^2
  +m_t^2)m_W^2]B_0^{Wbt}\}\nonumber,
\\&& \delta Z_W =\frac{g^2}{32\pi^2m_W^2}\{\frac{(m_b^2 -m_t^2)^2}
  {m_W^2}(B_0^{0bt} -B_0^{Wbt}) +[(m_b^2 -m_t^2)^2
  \nonumber\\ && \hspace{0.9cm} +(m_b^2 +m_t^2)m_W^2]B_0^{'Wbt}\}\nonumber,
\\&& \delta m_Z^2 =\frac{g^2s_W^2}{18c_W^2\pi^2}[\frac{m_b^2}{2}
  (3 -2s_W^2)(B_0^{Zbb} +B_0^{0bb}) -m_t^2(3 -4s_W^2)(B_0^{Ztt}
  -B_0^{0tt})]
  \nonumber\\ && \hspace{0.9cm} +\frac{g^2}{32c_W^2\pi^2}[m_b^2(B_0^{Zbb}
  -2B_0^{0bb}) -m_t^2(B_0^{Ztt} + 2 B_0^{0tt})]\nonumber,
\\&& \delta Z_{H^+} =\frac{3}{16\pi^2}[2(h_t^2\beta_{11}^2
  +h_b^2\beta_{21}^2)(B_1^{H^+bt} +m_b^2B_0^{'H^+bt}
  +m_{H^+}^2B_1^{'H^+bt})
  \nonumber\\ && \hspace{1.1cm} -4h_bh_tm_bm_t\beta_{11}\beta_{21}
  B_0^{'H^+bt} +\sum_{i,j,i',j'} (\theta_{ii'}^b)^2(\theta_{jj'}^t)^2(h_b
  \Theta_{i'j'1}^5 +h_t\Theta_{i'j'1}^6)^2B_0^{'H^+\tilde b_i
  \tilde t_j}]\nonumber,
\\&& \delta Z_h =\frac{3}{16\pi^2}\{2h_t^2\alpha_{11}^2
  (B_1^{htt} +2m_t^2B_0^{'htt} +m_h^2B_1^{'htt})
  +2h_b^2\alpha_{21}^2(B_1^{hbb}
  \nonumber\\ && \hspace{1.1cm} +2m_b^2B_0^{'hbb}
  +m_h^2B_1^{'hbb}) +\sum_{i,j,i',j'} [(h_t\theta_{ii'}^t\theta_{jj'}^t
  \Theta_{i'j'1}^1)^2B_0^{'h\tilde t_i\tilde t_j}
  \nonumber\\ && \hspace{1.1cm} +(h_b\theta_{ii'}^b \theta_{jj'}^b
  \Theta_{i'j'1}^2)^2 B_0^{'h\tilde b_i\tilde b_j}]\}\nonumber,
\\&&Z_{H^+W}=\frac{-3g}{16\sqrt{2}\pi^2m_{H^+}^2m_W^2} [
(h_tm_t\beta_{11}+h_bm_b\beta_{12})
((m_b^2-m_t^2)(B_0^{0bt}-B_0^{H^+bt})-m_{H
^+}^2B_0^{H^+bt})\nonumber\\
&&\hspace{1.1cm}+\sum_{i,j,i',j'}\theta^b_{i1}\theta^b_{ii'}\theta^t_{j1}\theta^t_{jj'}
(h_b\Theta^5_{i'j'1}+h_t\Theta^6_{i'j'1}) (m_{\tilde
t_j}^2-m_{\tilde b_i}^2) (B_0^{0\tilde b_i \tilde t_j}
-B_0^{H^+\tilde b_i\tilde t_j})],\nonumber\\
&&Z_{Hh}^{1/2}=\frac{3\alpha_{11}\alpha_{12}}{16\pi^2(m_{H}^2-m_{h}^2)}
[2m_b^2(1+B_0^{0bb}+2B_0^{hbb})-2m_t^2(1+B_0^{0tt}+2B_0^{htt})\nonumber\\
&&\hspace{1.1cm} -m_{h}^2(B_0^{hbb}-B_0^{htt})]\nonumber\\
&&\hspace{1.1cm}+\frac{3}{16\pi^2(m_{H}^2-m_{h}^2)}\sum_{i,j,i',j'}[
(h_b\theta^b_{ii'}\theta^b_{jj'})^2\Theta^2_{i'j'1}\Theta^2_{i'j'2}
B_0^{h\tilde b_i \tilde b_j}
+(h_t\theta^t_{ii'}\theta^t_{jj'})^2\Theta^1_{i'j'1}\Theta^1_{i'j'2}
B_0^{h\tilde t_i \tilde t_j} ]\nonumber\\
&&\hspace{1.1cm}-\frac{3\alpha_{11}\alpha_{12}}{16\pi^2(m_{H}^2-m_{h}^2)}\sum_i[
h_b^2m_{\tilde b_i}^2(1+B_0^{0\tilde b_i\tilde b_i})
+h_t^2m_{\tilde t_i}^2(1+B_0^{0\tilde t_i\tilde t_i})],\nonumber\\
\end{eqnarray}
  with
\begin{eqnarray}
&&
B_{n}^{ijk}=(-1)^n\{\frac{\Delta}{n+1}-\int_{0}^{1}dyy^{n}\ln{[\frac{m_i^{2}
y(y-1) +m_{j}^{2}(1-y)+m_{k}^{2}y}{\mu^{2}}]}\},
\\ && B_{n}^{'ijk}
=(-1)^n\int_{0}^{1}dy\frac{y^{n+1}(1-y)} {m_i^2 y(y-1)
+m_{j}^{2}(1-y) +m_{k}^{2}y}.
\end{eqnarray}
Here $h_b\equiv gm_b/\sqrt{2}m_W\cos\beta$ and $h_t\equiv
gm_t/\sqrt{2}m_W\sin\beta$ are the Yukawa couplings from the
bottom and top quarks, respectively. The notations
$\theta^t_{ij}$, $\theta^b_{ij}$, $\alpha_{ij}$, $\beta_{ij}$,
$\varphi_{ij}$ and $\Theta^n_{ijk}$ used in the above expressions
are defined in Appendix A.

Calculating the diagrams in Fig.1$(b)$-$(e)$, we can get the
explicit expressions of the vertex form factors $\Lambda_j
(j=1$-$4)$ as following:
\begin{eqnarray}
&&\Lambda_1=-\frac{3g h_b}{16 \pi^2} \sin\alpha\{
   h_b\sin\beta(B_0^{H^+bt}-B_1^{H^+tb})+[
   -h_tm_bm_t\cos\beta(C_0+2C_2)\nonumber\\&&\hspace{1.1cm}
   -2h_b\sin\beta C_{00}
   +h_bm_b^2\sin\beta(C_0-2C_2)
   -h_bm_{H^+}^2\sin\beta(C_{12}+4C_{22})\nonumber\\&&\hspace{1.1cm}
   +h_bm_{h}^2\sin\beta(C_{12}+2C_{22})
   +h_bm_{W}^2\sin\beta(C_1-3C_{12}\nonumber\\&&\hspace{1.1cm}
   -4C_{22})](m_W^2,m_{H^+}^2,m_h^2,m_b^2,m_t^2,m_b^2)\},\nonumber\\&&
\Lambda_2=\frac{3g h_t}{16 \pi^2} \cos\alpha \{
   h_t\cos\beta(-B_0^{H^+bt}+B_1^{H^+bt})+[
   h_bm_bm_t\sin\beta(C_0+2C_2)\nonumber\\&&\hspace{1.1cm}
   +2h_t\cos\beta C_{00}
   -h_tm_t^2\cos\beta(C_0-2C_2)
   +h_tm_{H^+}^2\cos\beta(C_{12}+4C_{22})\nonumber\\&&\hspace{1.1cm}
   +h_tm_{h}^2\cos\beta(C_{12}+2C_{22})
   -h_tm_{W}^2\cos\beta(C_1-3C_{12}\nonumber\\&&\hspace{1.1cm}
   -4C_{22})](m_W^2,m_{H^+}^2,m_h^2,m_t^2,m_b^2,m_t^2)\},\nonumber\\&&
\Lambda_3=\frac{3 g h_b}{8\sqrt{2} \pi^2} \sum_{ijk}\sum_{i'j'k'}
   \theta^b_{i1}\theta^b_{ii'} (\theta^t_{jj'})^2\theta^t_{k1}\theta^t_{kk'}\Theta^2_{i'j'1}
   (h_b\Theta^5_{j'k'1}+h_t\Theta^6_{j'k'1})\nonumber\\&&\hspace{1.1cm}
   \times C_2(m_W^2,m_{H^+}^2,m_h^2,m_{\tilde b_i}^2,m_{\tilde t_k}^2,m_{\tilde b_j}^2),\nonumber
\\&& \Lambda_4=-\frac{3 g h_t}{8\sqrt{2} \pi^2} \sum_{ijk}\sum_{i'j'k'}
  \theta^b_{k1}\theta^b_{kk'}
  \theta^t_{i1}\theta^t_{ii'}(\theta^t_{jj'})^2\Theta^1_{j'i'1}
  (h_b\Theta^5_{k'j'1}+h_t\Theta^6_{k'j'1})\nonumber\\&&\hspace{1.1cm}
  \times C_2(m_W^2,m_{H^+}^2,m_h^2,m_{\tilde t_i}^2,m_{\tilde b_k}^2,m_{\tilde
  t_j}^2),
\end{eqnarray}
where $C_i$ and $C_{ij}$ are the three-point Feynman
integrals[10].

The corresponding amplitude squared is
\begin{eqnarray}
|M_{ren}|^{2} =|M_0|^{2} +2Re(\delta M M_0^{\dag}).
\end{eqnarray}
The renormalized decay width is given by
\begin{eqnarray}
  \Gamma(H^+\rightarrow W^+h)&=&\Gamma_0(H^+\rightarrow
W^+h)+\delta\Gamma(H^+\rightarrow W^+h)\nonumber\\
  &=&\frac{1}{16\pi
m_{H^+}^3}\sqrt{(m_h^2-m_{H^+}^2-m_W^2)^2-4m_{H^+}^2m_W^2}
|M_{ren}|^{2}.
\end{eqnarray}
The tree-level branching ratio of $H^+\rightarrow W^+h$ decay is
\begin{eqnarray}
B_0(H^+\rightarrow W^+h)=\Gamma_0(H^+\rightarrow
W^+h)/\Gamma_0(H^+),
\end{eqnarray}
where the tree-level total decay width $\Gamma_0(H^+)$ of the
charged Higgs boson is approximated by
\begin{eqnarray}
\Gamma_0(H^+)=\Gamma_0(H^+\rightarrow W^+h) +
\Gamma_0(H^+\rightarrow tb)+\Gamma_0(H^+\rightarrow
  cs)+\Gamma_0(H^+\rightarrow\tau\nu).
\end{eqnarray}
While calculating the one-loop branching ratio $B(H^+\rightarrow
W^+h)$, the QCD corrections[11] to the $H^+\rightarrow tb$ width,
$\delta \Gamma(H^+\rightarrow tb)$, are included into the total
width $\Gamma(H^+)$, but its leading EW corrections[12] are
neglected since they are much smaller than the QCD corrections:
\begin{eqnarray}
B(H^+\rightarrow W^+h)=\Gamma(H^+\rightarrow W^+h)/\Gamma(H^+)
\end{eqnarray}
with
\begin{eqnarray}
\Gamma(H^+)=\Gamma_0(H^+)+\delta\Gamma(H^+\rightarrow W^+h) +
\delta \Gamma(H^+\rightarrow  tb)
\end{eqnarray}

\vspace{.4cm}

\begin{center}{\Large 3. Numerical results and conclusion}\end{center}

We now present some numerical results for the SUSY EW corrections
to the  decay $H^+\rightarrow W^+h$. The SM input parameters in
our calculations were taken to be $\alpha_{ew}(m_Z)=1/128.8$,
$m_W=80.375$GeV and $m_Z=91.1867$GeV[13], and $m_t=175.6$GeV and
$m_b=4.7$GeV. Other parameters are determined as follows

(i)  The one-loop relations[14] between the Higgs boson masses
$M_{h,H,A,H^\mp}$ and the parameters $\alpha$ and $\beta$ in the
MSSM were used, and $m_{H^+}$ and $\beta$ were chosen as the two
independent input parameters. As explained in Re[9], there is a
small inconsistency in doing so since the parameters $\alpha$ and
$\beta$ of Ref[14] are not the ones defined by the conditions
given by Eq.(Eq(3)) of Ref[9]. Neverthless, this difference would
only induce a higher order change[9].

(ii) For the parameters $m^2_{\tilde{Q},\tilde{U},\tilde{D}}$ and
$A_{t,b}$ in squark mass matrices
\begin{eqnarray}
M^2_{\tilde{q}} =\left(\begin{array}{cc} M_{LL}^2 & m_q M_{LR}\\
m_q M_{RL} & M_{RR}^2 \end{array} \right)
\end{eqnarray}
with
\begin{eqnarray}
&&M_{LL}^2 =m_{\tilde{Q}}^2 +m_q^2 +m_Z^2\cos 2\beta(I_q^{3L}
-e_q\sin^2\theta_W), \nonumber
\\&& M_{RR}^2 =m_{\tilde{U},\tilde{D}}^2 +m_q^2 +m_Z^2
\cos 2\beta e_q\sin^2\theta_W, \nonumber
\\&& M_{LR} =M_{RL} =\left(\begin{array}{ll} A_t -\mu\cot\beta &
(\tilde{q} =\tilde{t}) \\ A_b -\mu\tan\beta & (\tilde{q}
=\tilde{b}) \end{array} \right),
\end{eqnarray}
to simplify the calculation we assumed
$M_{\tilde{Q}}=M_{\tilde{U}} =M_{\tilde{D}}$ and $A_t=A_b$, and we
used $M_{\tilde{Q}}$, $A_t$ and $\mu$ as input parameters except
the numerical calculations as shown in Fig.5(a)and (b), where we
took $m_{\tilde t_1}$, $m_{\tilde b_1}$, $A_t=A_b$ and $\mu$ as
the input parameters.

Figure 3 shows the tree-level partial width are relatively large
for low values of $\tan\beta(=1.5,2)$ when $m_{H^+}>200GeV$.

In Figs.4(a) and (b) we present the SUSY EW corrections to the
tree-level decay widths and the braching radios as functions of
$m_{H^+}$ for different values of $\tan\beta$, respectively,
assuming $M_{\tilde Q}=M_{\tilde t}=M_{\tilde b}=300$GeV,
$A_t=A_b=300$GeV, and $\mu=-100$GeV. Figure 4(a) shows that the
relative corrections increase the partial width significantly at
low $\tan\beta$, which can exceed $40\%$ for $\tan\beta=1.5$ and
$30\%$ for $\tan\beta=2$, respectively. Fig.4(b) gives the
tree-level branching ratios and the branching ratios after
including $O(\alpha_{ew} m_{t(b)}^{2}/ m_{W}^{2})$ and
$O(\alpha_{ew} m_{t(b)}^4 / m_W^4)$ SUSY EW corrections for
$\tan\beta=1.5$ and $3$, respectively. From Fig.4(b), we see that
the branching ratios are enhanced by the corrections, especially
for $m_{H^+} > 300$ GeV, they can be increased by $50\%$. But the
corrections to the branching ratios decrease with an decrease of
$m_{H^+}$, especially for $m_{H^+}$ bellow $200$ GeV, they become
negligibly small.

In Fig.5(a) and (b) we assumed $M_{\tilde Q}=M_{\tilde t}$,
$A_t=A_b=500$GeV, $\mu=100$GeV, $m_{\tilde t_1}=170$GeV and
$m_{\tilde b_1}=200$GeV. Fig.5(a) shows the relative corrections
to the tree-level widths as a function of $m_{H^+}$ for $\tan\beta
= 1.5$, $2$, $6$, $10$ and $30$. Comparing Fig.5(a) with Fig.4(a),
we see that the relative corrections are enhanced by the
relatively light squark masses in the case of heavy charged Higgs
masses at low or high $\tan\beta$. Fig.5(a) also shows at low
$\tan\beta$ the corrections always increase the decay width ,
which can exceed $50\%$, while at high $\tan\beta$ the corrections
decrease the decay width significantly. There are dips at
$m_{H^+}=m_{\tilde t_1}+m_{\tilde b_1}=370$GeV due to the
threshold effects. As shown in Fig.4(b), Fig.5(b) shows the
branching ratios are enhanced by the the SUSY EW corrections for
$\tan\beta=1.5$ and $3$, especially when $m_{H^+} > 300$GeV.

In conclusion, we have calculated the $O(\alpha_{ew} m_{t(b)}^{2}
/ m_{W}^{2})$ and $O(\alpha_{ew} m_{t(b)}^4 / m_W^4)$ SUSY EW
corrections to the process $H^+\rightarrow W^+h$ in the MSSM. In
general, these corrections increase the decay widths and the
branching radios and are sensitive to both of $\tan\beta $ and the
mass of the charged Higgs boson. For $m_{H^+}>200$GeV at low
$\tan\beta(\leq 3)$, the corrections can increase the decay widths
and the branching radios more than  $20\%$ and $40\%$,
respectively.

\vspace{.5cm}

This work was supported in part by the National Natural Science
Foundation of China, the Doctoral Program Foundation of Higher
Education of China, and a grant from the State Commission of
Science and Technology of China. \vspace{.5cm}

\begin{center}{\large Appendix A} \end{center}
We present some notations used in this paper here. We introduce an
angle $\varphi=\beta-\alpha$, and for each angle $\alpha$,
$\beta$, $\varphi$, $\theta^t$ or $\theta^b$, we define
\begin{eqnarray*}
&&\alpha_{ij}=\left(\begin{array}{cc} \cos\alpha & \sin\alpha\\
-\sin\alpha & \cos\alpha\end{array} \right),
\beta_{ij}=\left(\begin{array}{cc} \cos\beta & \sin\beta\\
-\sin\beta & \cos\beta\end{array} \right),
\varphi_{ij}=\left(\begin{array}{cc} \cos\varphi & \sin\varphi\\
-\sin\varphi & \cos\varphi\end{array} \right),\nonumber\\
&&\theta_{ij}^t=\left(\begin{array}{cc} \cos\theta^t &
\sin\theta^t\\ -\sin\theta^t & \cos\theta^t\end{array} \right),
\theta_{ij}^b=\left(\begin{array}{cc} \cos\theta^b &
\sin\theta^b\\ -\sin\theta^b & \cos\theta^b\end{array} \right)
\end{eqnarray*}
We define six matrix $\Theta^i_{jkl} (i=1-6)$ for the couplings
between squarks and Higgses:
\begin{eqnarray*}
&\Theta^1_{ij1}=&\frac{1}{\sqrt{2}} \left(\begin{array}{cc}
2m_t\cos\alpha &A_t\cos\alpha+\mu\sin\alpha\\
A_t\cos\alpha+\mu\sin\alpha& 2m_t\cos\alpha\end{array}
\right)\nonumber\\ &\Theta^1_{ij2}=&\frac{1}{\sqrt{2}}
\left(\begin{array}{cc} 2m_t\sin\alpha
&A_t\sin\alpha-\mu\cos\alpha\\ A_t\sin\alpha-\mu\cos\alpha&
2m_t\sin\alpha\end{array} \right)\\
&\Theta^2_{ij1}=&\frac{-1}{\sqrt{2}} \left(\begin{array}{cc}
2m_b\sin\alpha &A_b\sin\alpha+\mu\cos\alpha\\
A_b\sin\alpha+\mu\cos\alpha&
2m_b\sin\alpha\end{array}\right)\nonumber\\
&\Theta^2_{ij2}=&\frac{1}{\sqrt{2}} \left(\begin{array}{cc}
2m_b\cos\alpha &A_b\cos\alpha-\mu\sin\alpha\\
A_b\cos\alpha-\mu\sin\alpha& 2m_b\cos\alpha\end{array} \right)\\
&\Theta^5_{ij1}=&\left(\begin{array}{cc} m_b\sin\beta &0\\
A_b\sin\beta+\mu\cos\beta&m_t\sin\beta
\end{array} \right)\nonumber\\
&\Theta^5_{ij2}=&\left(\begin{array}{cc} -m_b\cos\beta &0\\
-A_b\cos\beta+\mu\sin\beta&0
\end{array} \right)\\
&\Theta^6_{ij1}=&\left(\begin{array}{cc} m_t\cos\beta
&A_t\cos\beta+\mu\sin\beta\\ 0&m_b\cos\beta
\end{array} \right)\nonumber\\
&\Theta^6_{ij2}=&\left(\begin{array}{cc} m_t\sin\beta
&A_t\sin\beta-\mu\cos\beta\\ 0&0
\end{array} \right)
\end{eqnarray*}

\newpage
\baselineskip=0.25in {\LARGE References} \vspace{0.2cm}
\begin{itemize}
\item[{\rm[1]}] H.E. Haber and G.L. Kane, Phys. Rep. 117, 75(1985);
                J.F. Gunion and H.E. Haber, Nucl. Phys. B272, 1(1986).
\item[{\rm[2]}] CMS Technical Proposal. CERN/LHC94-43 LHCC/P1, December 1994.
\item[{\rm[3]}] CDF Collaboration, Phys. Rev. Lett. 79, 35(1997);
                D0 Collaboration, Phys. Rev.Lett. 82, 4975(1999).
\item[{\rm[4]}] V.Barger, R.J.N. Phillips and D.P. Roy, Phys.Lett. B324,
                236(1994);
                J.F. Gunion and S.Geer, preprint UCD-93-32, September 1993,
                hep-ph/9310333; J.F. Gunion, Phys.lett. B322, 125(1994);
                D.J. Miller, S.Moretti, D.P. Roy and W.J. Stirling,
                Phys.Rev. D61,
                055011(2000); S. Moretti and D.P. Roy, B470 209(1999).
\item[{\rm[5]}] K. Odagiri, preprint RAL-TR-1999-012, February 1999,
                hep-ph/9901432; S. Raychaudhuri and D.P. Roy,
                Phys.Rev. D53, 4902(1996).
\item[{\rm[6]}] R.Santos, A.Barroso, and L. Brucher, Phys. Lett. B391,
                429(1997).
\item[{\rm[7]}] Stefano Moretti, Phys.Lett. B481, 49(2000).
\item[{\rm[8]}] S. Sirlin, Phys. Rev. D22, 971 (1980);
                W. J. Marciano and A. Sirlin,{\sl ibid.} 22, 2695(1980);
                31, 213(E) (1985);
                A. Sirlin and W.J. Marciano, Nucl. Phys. B189, 442(1981);
                K.I. Aoki et.al., Prog. Theor. Phys. Suppl. 73, 1(1982).
\item[{\rm[9]}] A. Mendez and A. Pomarol, Phys.Lett. B279, 98(1992).
\item[{\rm[10]}] G.Passarino and M.Veltman, Nucl. Phys. B160,
                 151(1979); A.Axelrod, {\sl ibid.} B209, 349 (1982);
                 M.Clements {\sl et al.}, Phys. Rev. D27, 570
                 (1983); A.Denner, Fortschr. Phys. 41, 4 (1993); R.
                 Mertig {\sl et al.}, Comput. Phys. Commun. 64, 345
                 (1991).
\item[{\rm[11]}] C.S. Li and R.J. Oakes, Phys. Rev. D43, 855(1991);
                 A. Mendez and A. Pomarol, Phys.Lett. B252, 461(1990);
                 C.S. Li and J. M. Yang, Phys.Lett. B315, 367(1993);
                 Heinz Konig, Modern Phys. Lett. A10, 1113(1995);
                 A. Bartl, H. Eberl, K. Hidaka, T. Kon, W. Majerotto and Y.
                 Yamada, Phys. Lett. B378, 167(1996);
                 A. Djoual, M. Spira and P.M. Zerwas, Z. Phys. C70, 427(1996).
\item[{\rm[12]}] J. M. Yang and C.S. Li, Phys. Rev. D47, 2872(1993);
                 M. A. Diaz, Phys.Rev. D48, 2152(1993)
\item[{\rm[13]}] Particle Data Group, C.Caso {\it et al}, Eur.Phys.J. C3,
                 1(1998).
\item[{\rm[14]}] J.Gunion, A.Turski, Phys. Rev. D39, 2701(1989);
                 D40, 2333(1990); J.R.Espinosa, M.Quiros, Phys. Lett.
                 B266, 389(1991); M.Carena, M.Quiros, C.E.M.Wagner, Nucl. Phys.
                 B461, 407(1996).
\end{itemize}
\newpage

\newcounter{fig}
\section *{Figure Captions}
\begin{list}{{\bf FIG. \arabic{fig}}}{\usecounter{fig}}
\item Feynman diagrams contributing to supersymmetric
      electroweak corrections to $H^+\rightarrow W^+h$: $(a)$ is the
      tree-level diagram; $(b)-(e)$ are the one-loop diagrams.
\item Feynman diagrams contributing to the renormalization constants.
\item The tree-level $H^+\rightarrow W^+h$ decay widths.
\item $(a)$: The SUSY EW relative corrections to the $H^+\rightarrow W^+h$ decay width;
      $(b)$: The branching ratio of the $H^+\rightarrow W^+h$ decay, assuming
      $M_{\tilde Q}=M_{\tilde t}=M_{\tilde b}=300$GeV, $A_t=A_b=300$GeV,
      and $\mu=-100$GeV. Logarithmic coordinate is used in (b).
\item $(a)$: The SUSY EW relative corrections to the $H^+\rightarrow W^+h$ decay width;
      $(b)$: The branching ratio of the $H^+\rightarrow W^+h$ decay, assuming
      $M_{\tilde Q}=M_{\tilde t}$, $A_t=A_b=500$GeV, $\mu=100$GeV,
      $m_{\tilde t_1}=170$GeV and $m_{\tilde b_1}=200$GeV. Logarithmic coordinate is used in
      (b).
\end{list}
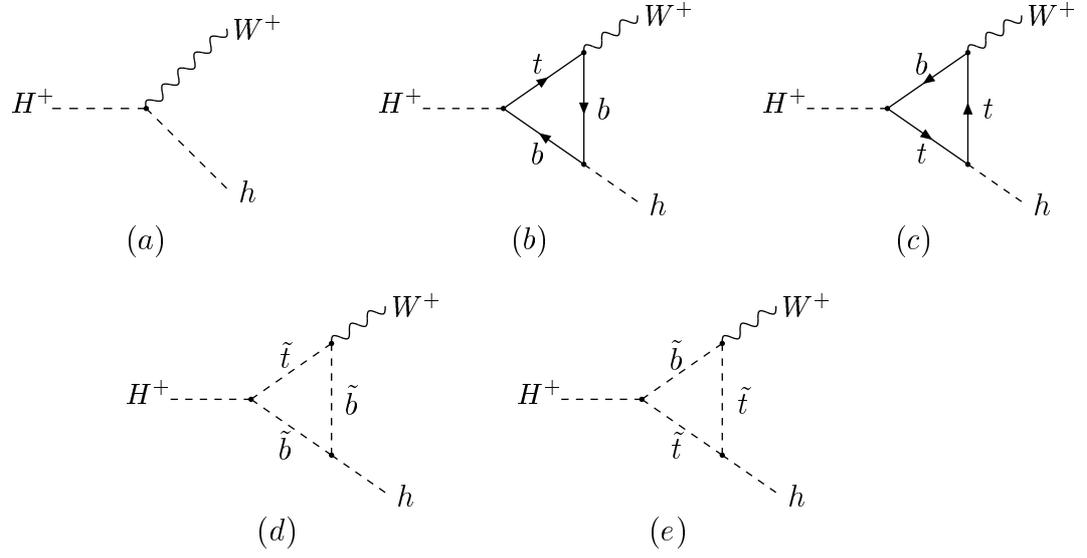
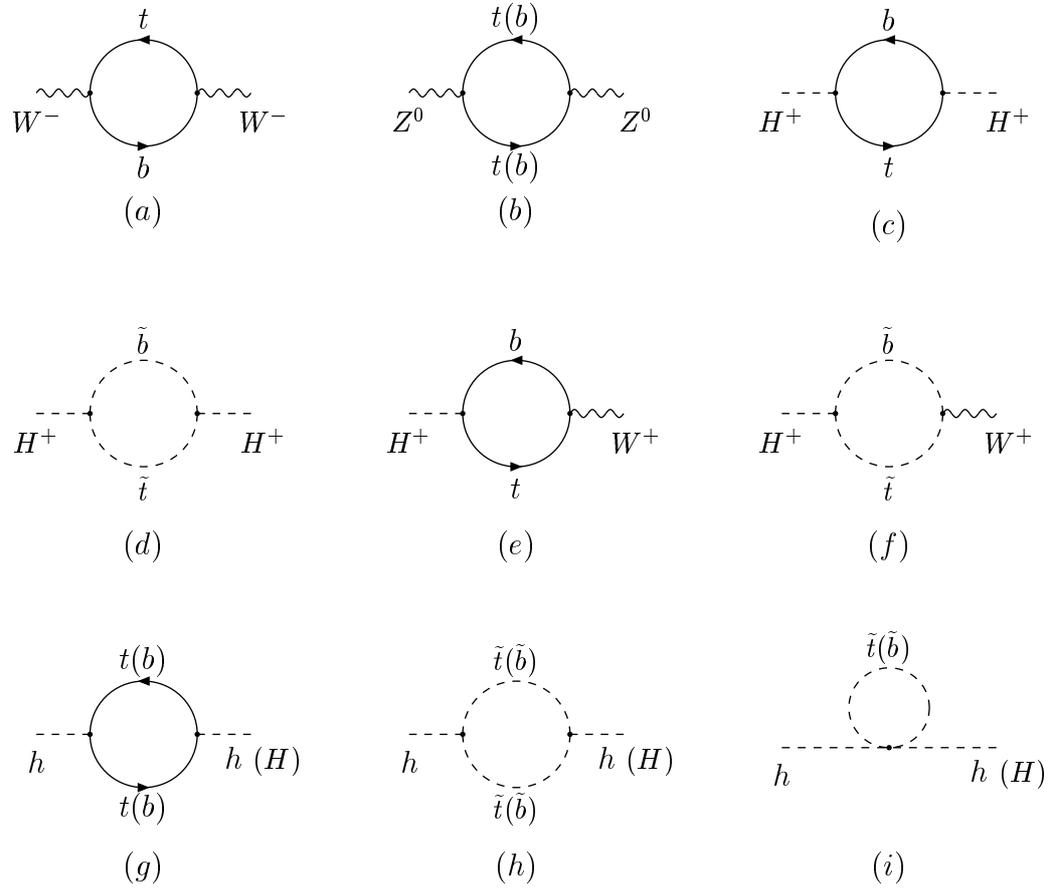
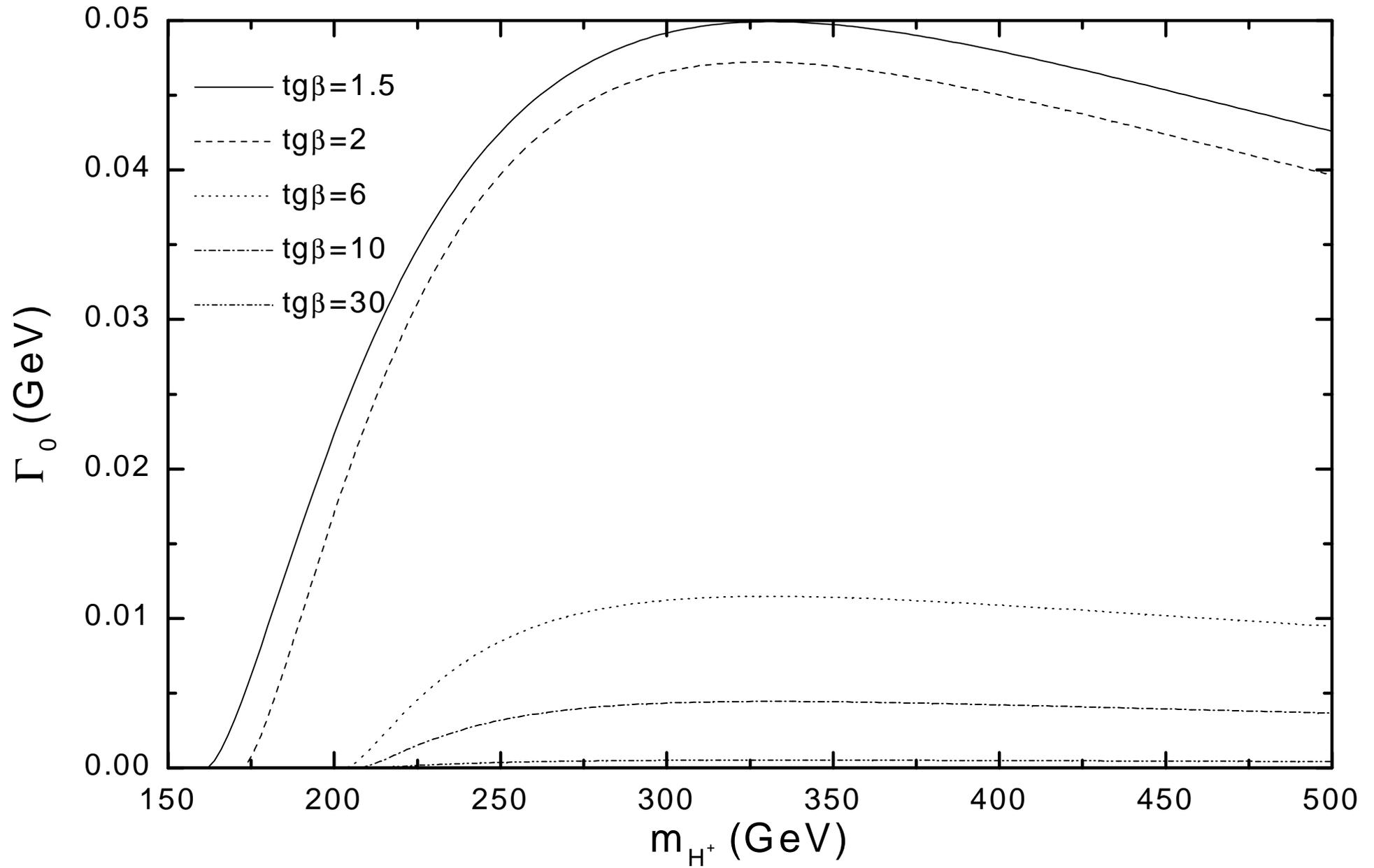
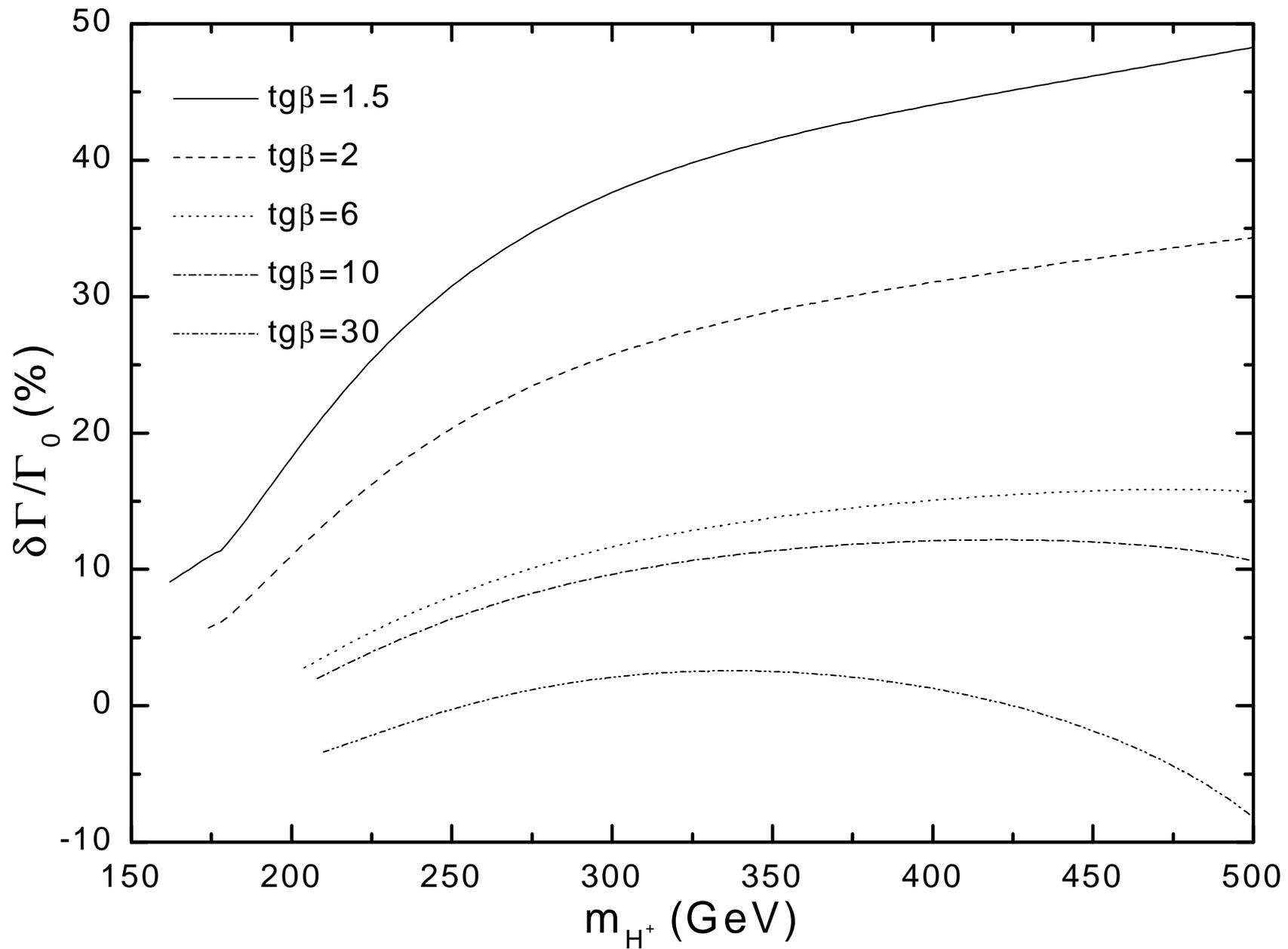
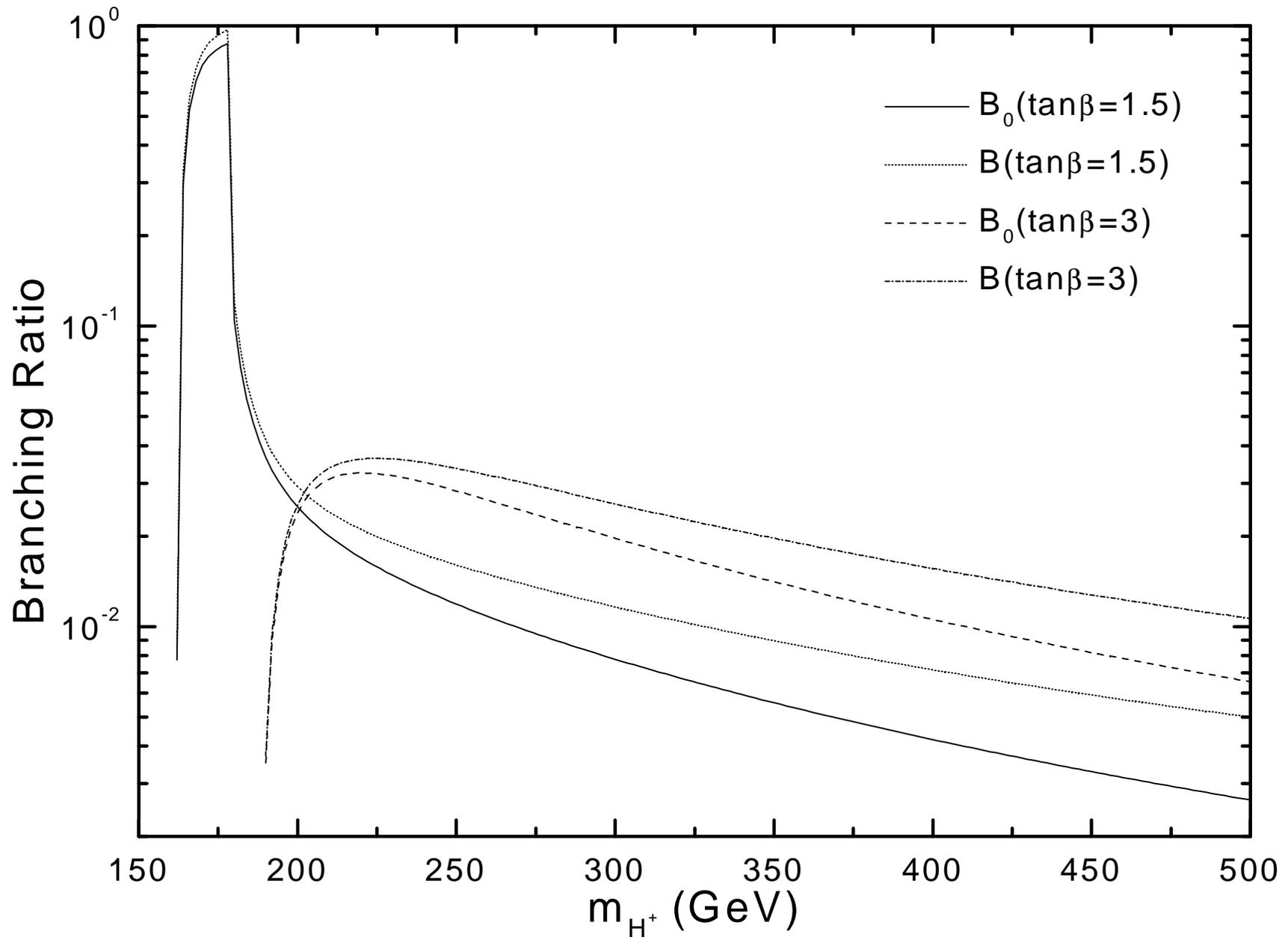
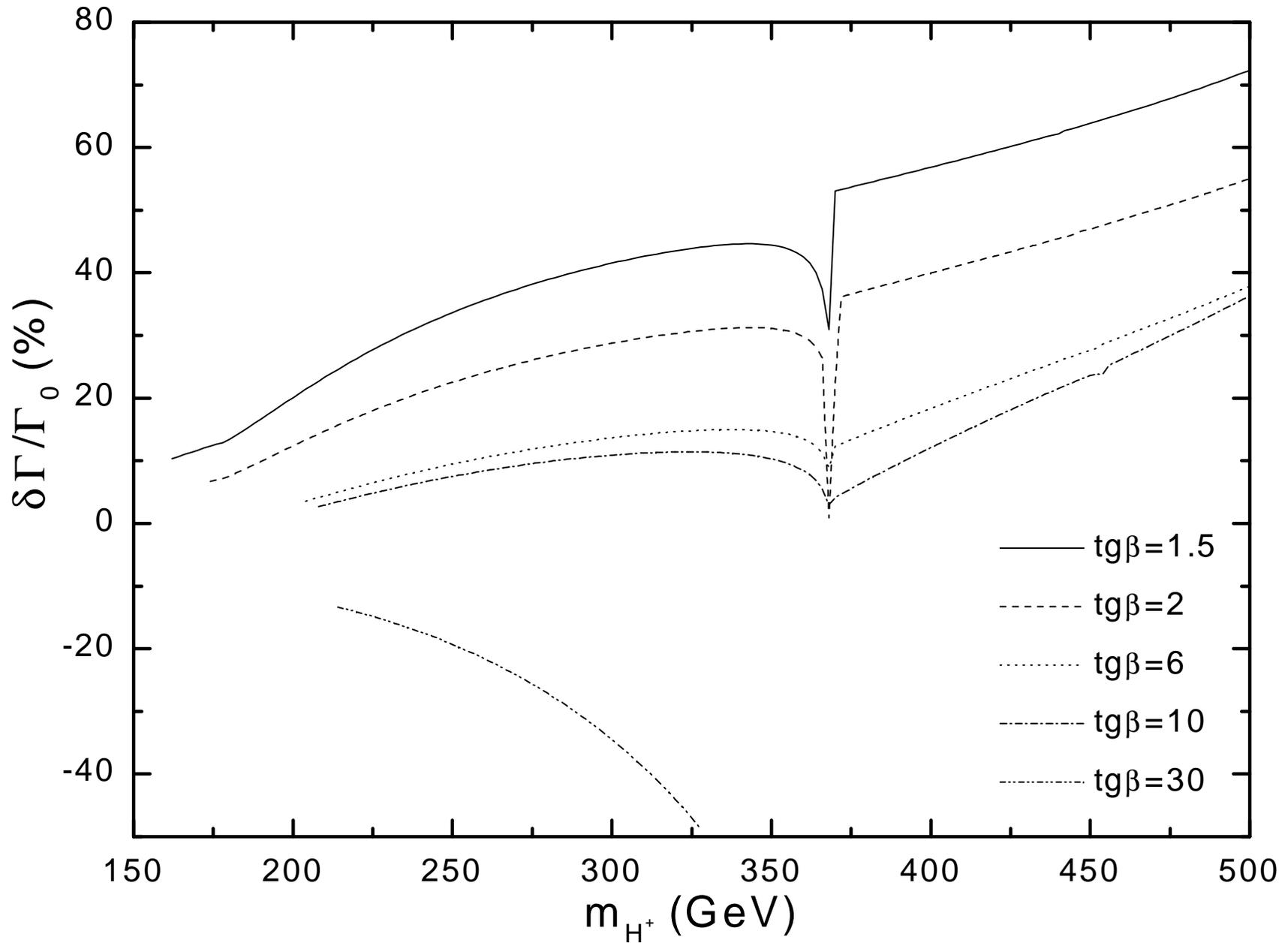
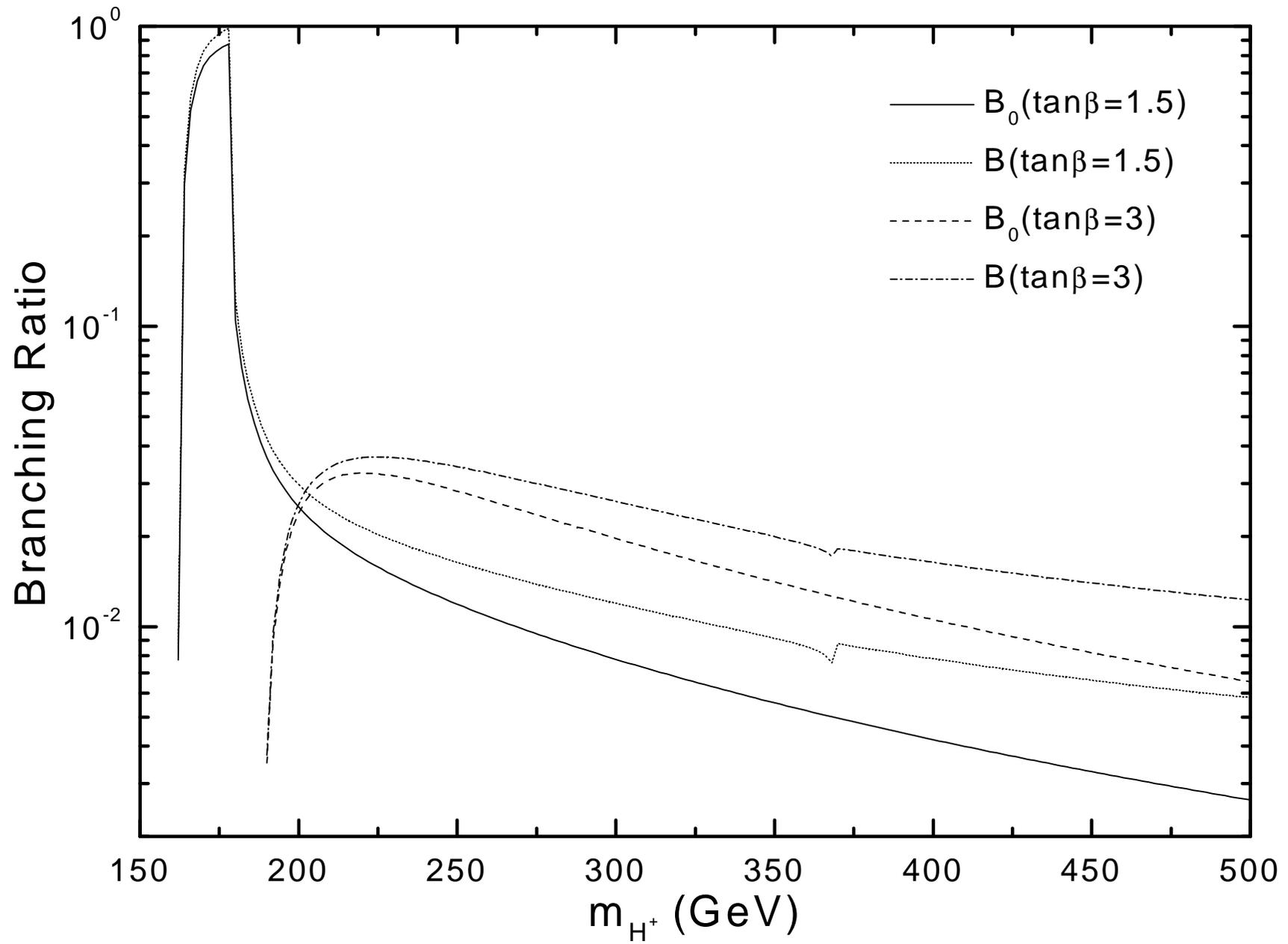
\end{document}